\documentclass[sigconf]{acmart}

\renewcommand\footnotetextcopyrightpermission[1]{} 
\AtBeginDocument{%
  \providecommand\BibTeX{{%
    \normalfont B\kern-0.5em{\scshape i\kern-0.25em b}\kern-0.8em\TeX}}}

\begin{document}

\title{XCloud: Design and Implementation of AI Cloud Platform with RESTful API Service}

\author{Lu Xu}
\email{xulu0620@gmail.com}
\author{Yating Wang}
\email{yt.one93@gmail.com}

\renewcommand{\shortauthors}{Lu Xu and Yating Wang}

\begin{abstract}
  In recent years, artificial intelligence (AI) has aroused much attention among both industrial and academic areas. However, building and maintaining efficient AI systems are quite difficult for many small business companies and researchers if they are not familiar with machine learning and AI. In this paper, we first evaluate the difficulties and challenges in building AI systems. Then an cloud platform termed \textit{XCloud}, which provides several common AI services in form of RESTful APIs, is constructed. Technical details are discussed in Section~\ref{sec:xcloud}. This project is released as open-source software and can be easily accessed for late research. Code is available at \url{https://github.com/lucasxlu/XCloud.git}.
\end{abstract}

%

\keywords{deep learning, cloud computing, computer vision, machine learning, artificial intelligence}

\maketitle

\section{Introduction}
\label{sec:intro}
Recent years have witnessed many breakthroughs in AI~\cite{he2016deep,lecun2015deep,silver2016mastering}, especially computer vision~\cite{krizhevsky2012imagenet}, speech recognition~\cite{amodei2016deep} and natural language processing~\cite{johnson2017google}. Deep learning models have surpassed human on many fields, such as image recognition~\cite{he2015delving} and skin cancer diagnosis~\cite{esteva2017dermatologist}. Face recognition has been widely used among smart phones (such as iPhone X FaceID~\footnote{\url{https://support.apple.com/en-us/HT208109}}) and security entrance. Recommendation system (such as Alibaba, Amazon and ByteDance) helps people easily find information they want. Visual search system allows us to easily get products by just taking a picture with cellphone~\cite{zhang2018visual,yang2017visual}.

\begin{figure}[tphb]
	\caption{Pipeline of building production-level AI service}
	\label{fig:pipeline}
	\includegraphics[width=0.47\textwidth]{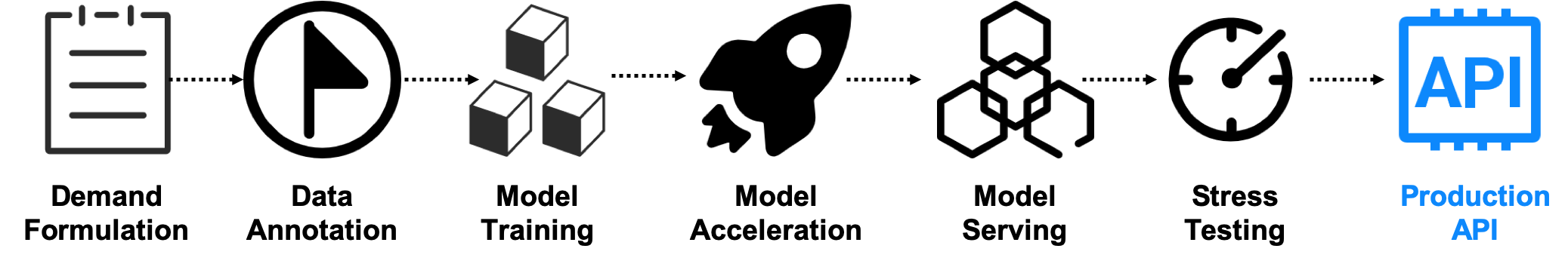}
\end{figure}

However, building an effective AI system is quite challenging~\cite{sculley2015hidden}. Firstly, the developers should collect, clean and annotate raw data to ensure a satisfactory performance, which is quite time-consuming and takes lots of money and energy. Secondly, experts in machine learning should formulate the problems and develop corresponding computational models. Thirdly, computer programmars should train models, fine-tune hyper-parameters, and develop SDK or API for later usage. Bad case analysis is also required if the performance of baseline model is far from satifaction. Last but not least, the above procedure should be iterated again and again to meet the rapid change of requirements (see Figure~\ref{fig:pipeline}). The whole development procedure may fail if any step mentioned above fails.

Facing so many difficulties, cloud services (such as Amazon Web Service (AWS)~\footnote{\url{https://aws.amazon.com/}}, Google Cloud~\footnote{\url{https://cloud.google.com/}}, AliYun~\footnote{\url{https://www.aliyun.com/}} and Baidu Yun~\footnote{\url{https://cloud.baidu.com/}}) are getting increasingly popular among market. Nevertheless, these platforms are developed for commercial production. Researchers only have limited access to existing APIs, and cannot know the inner design architecture of the systems. So it is difficult for researchers to bridge the gap between research models and production applications.

Aiming at solving problems mentioned above. In this paper, we construct an AI cloud platform termed \textit{EXtensive Cloud (XCloud)} with common recognition abilities for both research and production fields. \textit{XCloud} is freely accessible and open-sourced on github~\footnote{\url{https://github.com/lucasxlu/XCloud.git}} to help researchers build production application with their proposed models.

\section{XCloud}
\label{sec:xcloud}
\begin{figure}[htbp]
	\caption{Architecture of XCloud}
	\label{fig:arch}
	\includegraphics[width=0.5\textwidth]{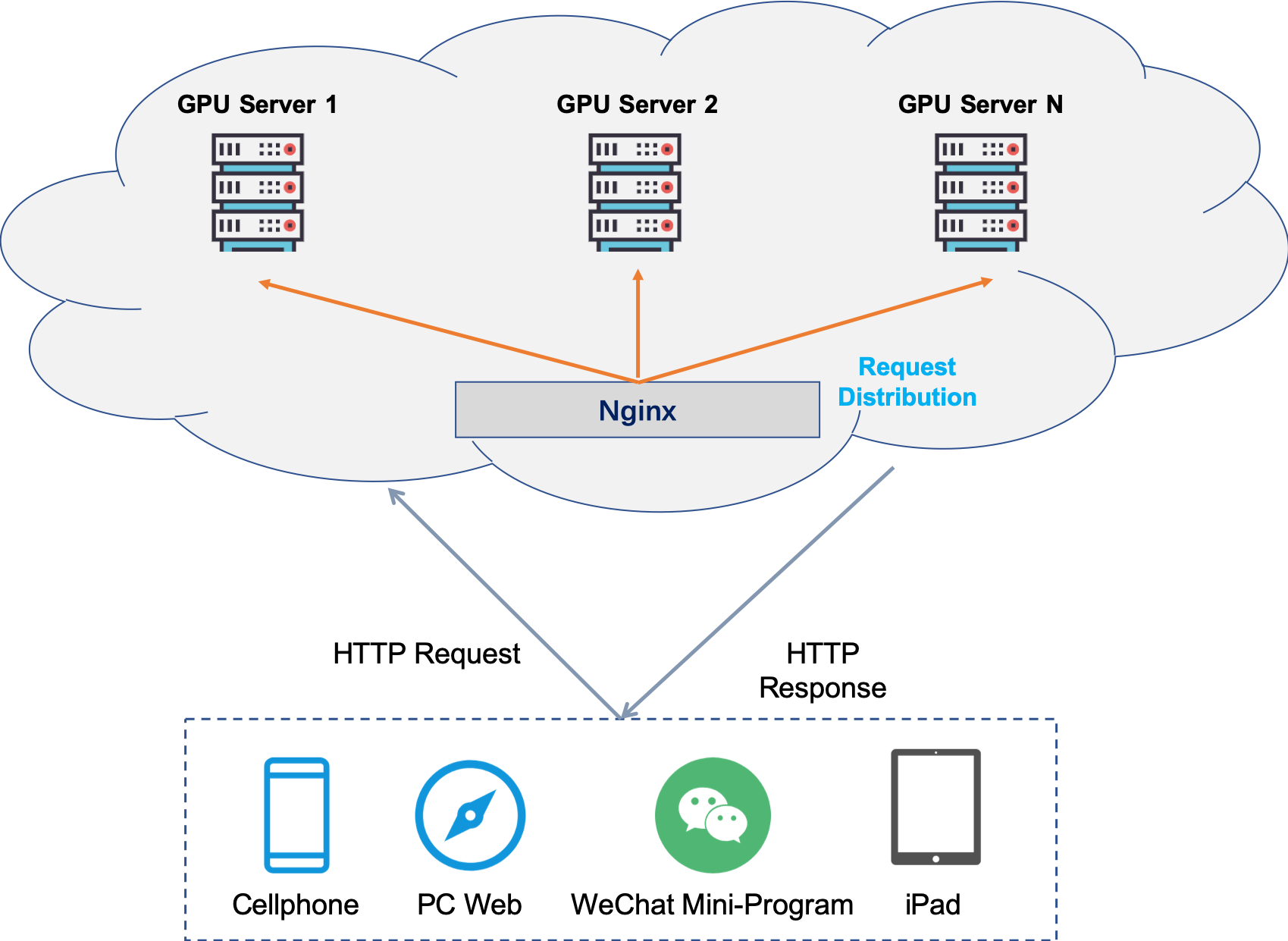}
\end{figure}

\begin{figure}[htbp]
	\caption{Framework of XCloud}
	\label{fig:framework}
	\includegraphics[width=0.5\textwidth]{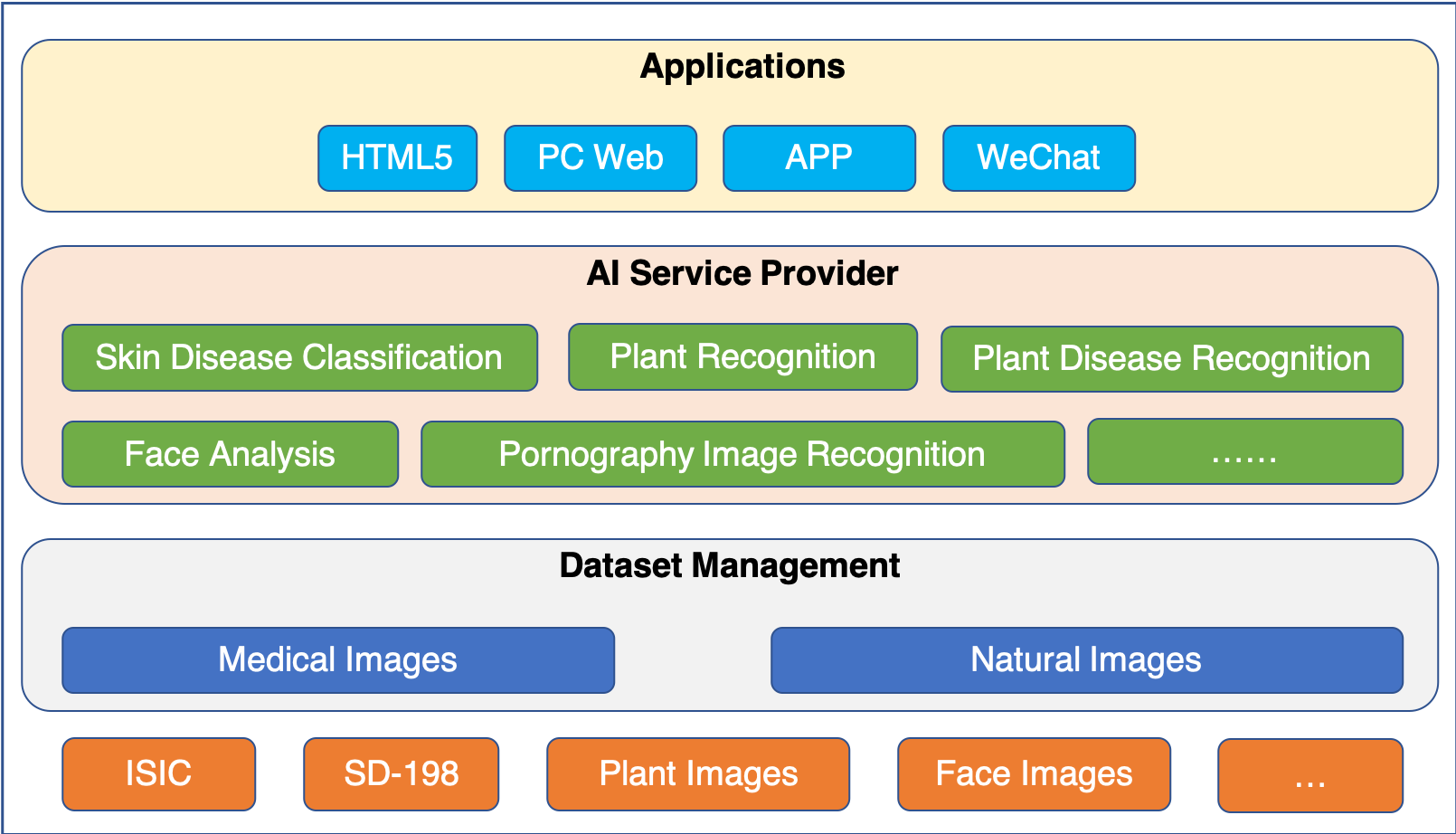}
\end{figure}

In this section, we will give a detailed description about the design and implementation of \textit{XCloud}. \textit{XCloud} is implemented based on PyTorch~\cite{paszke2019pytorch} and Django~\footnote{\url{https://www.djangoproject.com}}. The development of machine learning models are derived from published models~\cite{he2016deep,huang2017densely,xu2018crnet,xu2019hierarchical,xu2019data}, which is beyond the scope of this paper. The architecture of \textit{XCloud} is shown in Figure~\ref{fig:arch}. Users can upload image and trigger relevant JavaScript code, the controller of \textit{XCloud} receive HTTP request and call corresponding recognition APIs with the uploaded image as input. Then \textit{XCloud} will return recognition results in form of JSON. By leveraging RESTful APIs, the developers can easily integrate existing AI services into any type of terminals (such as PC web, android/iOS APPs and WeChat mini program). The overall framework of \textit{XCloud} is shown in Figure~\ref{fig:framework}.

\subsection{Services}
\textit{XCloud} is composed of 4 modules, namely, computer vision (CV), data mining (DM) and research (R). We will briefly introduce the following services by module.

\subsubsection{Computer Vision}
In CV module, we implement and train serveral models to solve the following common vision problems.
\begin{itemize}
	\item \textbf{Plants recognition} is popular among plant enthusiasts and botanists. It can be treated as a fine-grained visual classification problem, since a bunch of samples of different categories have quite similar appearance. We train ResNet18~\cite{he2016deep} to recognize over 998 plants.
	\item \textbf{Plant disease recognition} can provide efficient and effective tools in intelligent agriculture. Farmers can know disease category and take relevant measures to avoid huge loss. ResNet50~\cite{he2016deep} is trained to recognize over 60 plant diseases.
	\item \textbf{Face analysis} model can predict serveral facial attributes from a given portrait image. We take HMTNet~\cite{xu2019hierarchical} as computational backbone model. HMTNet is a multi-task deep model with fully convolutional architecture, which can predict facial beauty score, gender and race simultaneously from a unique model. Details can be found from~\cite{xu2019hierarchical}.
	\item \textbf{Food recognition} is popular among health-diet keepers and is widely used in \textit{New Ratailing} fields. DenseNet169~\cite{huang2017densely} is adopted to train food recognition model.
	\item \textbf{Skin lesion analysis} gains increased attention in medical AI areas. We train DenseNet121~\cite{huang2017densely} to recognize 198 common skin diseases.
	\item \textbf{Pornography image recognition} models provide helpful tools to filter sensitive images on Internet. We also integrate this feature into \textit{XCloud}. We train DenseNet121~\cite{huang2017densely} to recognize pornography images.
	\item \textbf{Garbage Classification} has been a hot topic in China recently~\footnote{\url{http://www.xinhuanet.com/english/2019-07/03/c_138195992.htm}}, it is an environment-friendly behavior. However, the majority of the people cannot tell different garbage apart. By leveraging computer vision and image recognition technology, we can easily classify diverse garbage. The dataset is collected from HUAWEI Cloud~\footnote{\url{https://developer.huaweicloud.com/competition/competitions/1000007620/introduction}}. We split 20\% of the images as test set, and the remaining as training set. We train ResNet152~\cite{he2016deep} with 90.12\% accuracy on this dataset.
	\item  \textbf{Insect Pet Recognition} plays a vital part in intelligent agriculture, we train DenseNet121~\cite{huang2017densely} on IP102 dataset~\cite{wu2019ip102} with 61.06\% accuracy, which is better than Wu et al.~\cite{wu2019ip102} with an improvement of 10.6\%.
\end{itemize}

%
%
%

\subsubsection{Data Mining}
In data mining module, we provide useful toolkit~\cite{xu2019data} related to an emerging research topic--\textbf{online knowledge quality evaluation} (like Zhihu Live~\footnote{\url{https://www.zhihu.com/lives/}}). This API will automatically calculate Zhihu Live's score within a range of 0 to 5, which can provide useful information for customers.

\subsubsection{Research}
In this module, we provide the source code for training and test machine learning models mentioned above. Researchers can use the code provided to train their own models. Furthermore, we also reimplement several models (such as image quality assessment~\cite{kang2014convolutional,bosse2016deep,talebi2018nima,kang2015simultaneous}, facial beauty analysis~\cite{xu2018crnet, xu2019hierarchical}, image retrieval~\cite{Liu_2017_CVPR,wen2016discriminative}, etc.) in computer vision, which makes it easy for users to integrate these features into XCloud APIs.

\subsection{Performance Metric}
The performance of the above models are listed in Table~\ref{tab:performance}. We adopt \textit{accuracy} as the performance metric to evaluate classification services (such as plant recognition, plant disease recognition, food recognition, skin lesion analysis and pornography image recognition), and \textit{Pearson Correlation (PC)} is utilized as the metric in facial beauty prediction task. Mean Absolute Error (MAE) is adopted as the metric in ZhihuLive quality evaluation task.

\begin{equation}
PC = \frac{\sum_{i=1}^{n}(x_i - \bar{x})(y_i - \bar{y})}{\sqrt{\sum_{i=1}^{n}(x_i-\bar{x})^2} \sqrt{\sum_{i=1}^{n}(y_i-\bar{y})^2}}
\end{equation}

\begin{equation}
MAE = \frac{1}{n}\sum_{i=1}^{n}|x_i - y_i|
\end{equation}

where $x_i$ and $y_i$ represent predicted score and groundtruth score, respectively. $n$ denotes the number of data samples. $\bar{x}$ and $\bar{y}$ stand for the mean of $x$ and $y$, respectively. A larger PC value represents better performance of the computational model.

\begin{table*}[htbp]
	\caption{Performance of Computational Models on Relevant Datasets}
	\label{tab:performance}
	\centering
	\begin{tabular}{ccccc}
		\hline
		\textbf{Service} & \textbf{Model} & \textbf{Dataset} & \textbf{Performance} & \textbf{Result} \\
		\hline \hline
		Plant Recognition & ResNet18~\cite{he2016deep} & FGVC5 Flowers~\footnote{\url{https://sites.google.com/view/fgvc5/competitions/fgvcx/flowers}} & Acc=0.8909 & Plant category and confidence \\
		Plant Disease Recognition & ResNet50~\cite{he2016deep} & PDD2018 Challenge~\footnote{\url{https://challenger.ai/dataset/pdd2018}} & Acc=0.8700 & Plant disease category and confidence \\
		Face Analysis & HMTNet~\cite{xu2019hierarchical} & SCUT-FBP5500~\cite{liang2018scut} & PC=0.8783 & Facial beauty score within $[1, 5]$ \\
		Food Recognition & DenseNet161~\cite{huang2017densely} & iFood~\footnote{\url{https://sites.google.com/view/fgvc5/competitions/fgvcx/ifood}} & Acc=0.6689 & Food category and confidence  \\
		Garbage Classification & ResNet152~\cite{he2016deep}  & HUAWEI Cloud & Acc=0.9012 & Garbage category and confidence \\
		Insect Pet Recognition & DenseNet121~\cite{huang2017densely} & IP102~\cite{wu2019ip102} & Acc=0.6106 & Insect pet category and confidence \\
		Skin Disease Recognition & DenseNet121~\cite{huang2017densely} & SD198~\cite{sun2016benchmark} & Acc=0.6455 & Skin disease category and confidence \\
		Porn Image Recognition & DenseNet121~\cite{huang2017densely} & nsfw\_data\_scraper~\footnote{\url{https://github.com/alexkimxyz/nsfw_data_scraper.git}} & Acc=0.9313 & Image category and confidence \\
		Zhihu Live Rating & MTNet~\cite{xu2019data} & ZhihuLiveDB~\cite{xu2019data} & MAE=0.2250 & Zhihu Live score within $[0, 5]$ \\
		\hline
	\end{tabular}
\end{table*}

\subsection{Design of RESTful API}
Encapsulating RESTful APIs is regarded as standard in building cloud platform. With RESTful APIs, related services can be easily integrated into terminal devices such as PC web, WeChat mini program, android/iOS APPs, and HTML5, without considering compatibility problems. The RESTful APIs provided are listed in Table~\ref{tab:rest_api}.

\begin{table*}[htbp]
	\caption{Definition of RESTful API}
	\label{tab:rest_api}
	\centering
	\begin{tabular}{cccc}
		\hline
		\textbf{API} & \textbf{Description} & \textbf{HTTP Methods} & \textbf{Param}\\\hline \hline
		cv/mcloud/skin & skin disease recognition & POST & imgraw/imgurl \\
		cv/fbp & facial beauty prediction & POST & imgraw/imgurl \\
		cv/nsfw & pornography image recognition & POST & imgraw/imgurl \\
		cv/pdr & plant disease recognition & POST & imgraw/imgurl \\
		cv/food & food recognition & POST & imgraw/imgurl \\
		cv/plant & plant recognition & POST & imgraw/imgurl \\
		cv/facesearch & face retrieval & POST & imgraw/imgurl \\
		dm/zhihuliveeval & Zhihu Live rating & GET & Zhihu Live ID \\
		\hline
	\end{tabular}
\end{table*}

\subsection{Backend Support}
The backend of \textit{XCloud} is developed based on Django~\footnote{\url{https://www.djangoproject.com/}}. We follow the \textit{MVC}~\cite{leff2001web} design pattern which represents that the view, controller and model are separately developed and can be easily extended in later development work. In order to record user information produced on \textit{XCloud}, we construct 2 relational tables in MySQL which is listed in Table~\ref{tab:db_api} and Table~\ref{tab:db_user}, to store relevant information.

\begin{table}[htbp]
	\caption{API calling details table. The primary key is decorated with underline.}
	\label{tab:db_api}
	\centering
	\begin{tabular}{cccc}
		\hline
		\textbf{Attribute} & \textbf{Type} & \textbf{Length} & \textbf{Is Null?} \\
		\hline \hline
		\underline{username} & varchar & 16 & False \\
		api\_name & varchar & 20 &  False \\
		api\_elapse & float & 10 & False \\
		api\_call\_datetime & datetime & - & False \\
		terminal\_type & int & 3 & False \\
		img\_path & varchar & 100 & False \\
		\hline
	\end{tabular}
\end{table}

\begin{table}[htbp]
	\caption{User information table. The primary key is decorated with underline.}
	\label{tab:db_user}
	\centering
	\begin{tabular}{cccc}
		\hline
		\textbf{Attribute} & \textbf{Type} & \textbf{Length} & \textbf{Is Null?} \\\hline\hline
		\underline{username} & varchar & 16 & False \\
		register\_datetime & datetime & - & False \\
		register\_type & int & 11 &  False \\
		user\_organization & varchar & 100 & False \\
		email & varchar & 50 & False \\
		userkey & varchar & 20 & False \\
		password & varchar & 12 & False \\
		\hline
	\end{tabular}
\end{table}

In addition, we also provide simple and easy-to-use script to convert original PyTorch models to TensorRT~\footnote{\url{https://developer.nvidia.com/tensorrt}} models for faster inference. TensorRT is a platform for high-performance deep learning inference. It includes a deep learning inference optimizer and runtime that delivers low latency and high-throughput for deep learning inference applications. With TensorRT, we are able to run DenseNet169~\cite{huang2017densely} with 97.63 FPS on two 2080TI GPUs, which is significantly faster than its counterpart PyTorch naive inference engine (29.45 FPS).

\subsection{Extensibility}
As shown by the name of XCloud (EXtensive Cloud), it is also quite easy to integrate new abilities. Apart from using existing AI technology provided by \textit{XCloud}, developers can also easily build their own AI applications by referring to the model training code contained in research module~\footnote{\url{https://github.com/lucasxlu/XCloud/tree/master/research}}. Hence, the developers only need to prepare and clean dataset. After training your own models, your AI interface is automatically integrated into \textit{XCloud} by just writing a new controller class and adding a new Django view.

\subsection{API Stress Testing}
The performance and stability play key roles in production-level service. In order to ensure the stability of \textit{XCloud}, Nginx~\footnote{\url{http://nginx.org/}} is adopted for load balancing. In addition, we use JMeter~\footnote{\url{https://jmeter.apache.org/}} to test all APIs provided by \textit{XCloud}. The results of stress testing can be found in Table~\ref{tab:test}.

\begin{table}[htbp]
	\caption{Stress Testing Results on NVIDIA 2080TI GPU}
	\label{tab:test}
	\centering
	\begin{tabular}{cccc}
		\hline
		\textbf{API} & \textbf{AVG\_LATENCY (ms)} & \textbf{P99 (ms)} & \textbf{ERROR}\\\hline \hline
		cv/mcloud/skin & 16 & 20 & 0 \\
		cv/fbp & 25 & 36 & 0 \\
		cv/nsfw & 16 & 21 & 0 \\
		cv/pdr & 16 & 23 & 0 \\
		cv/food & 17 & 23 & 0 \\
		cv/plant & 18 & 25 & 0 \\
		dm/zhihuliveeval & 5 & 8 & 0 \\
		\hline
	\end{tabular}
\end{table}

\begin{figure}[htbp]
	\caption{Homepage of \textit{XCloud}}
	\label{fig:index}
	\includegraphics[width=0.5\textwidth]{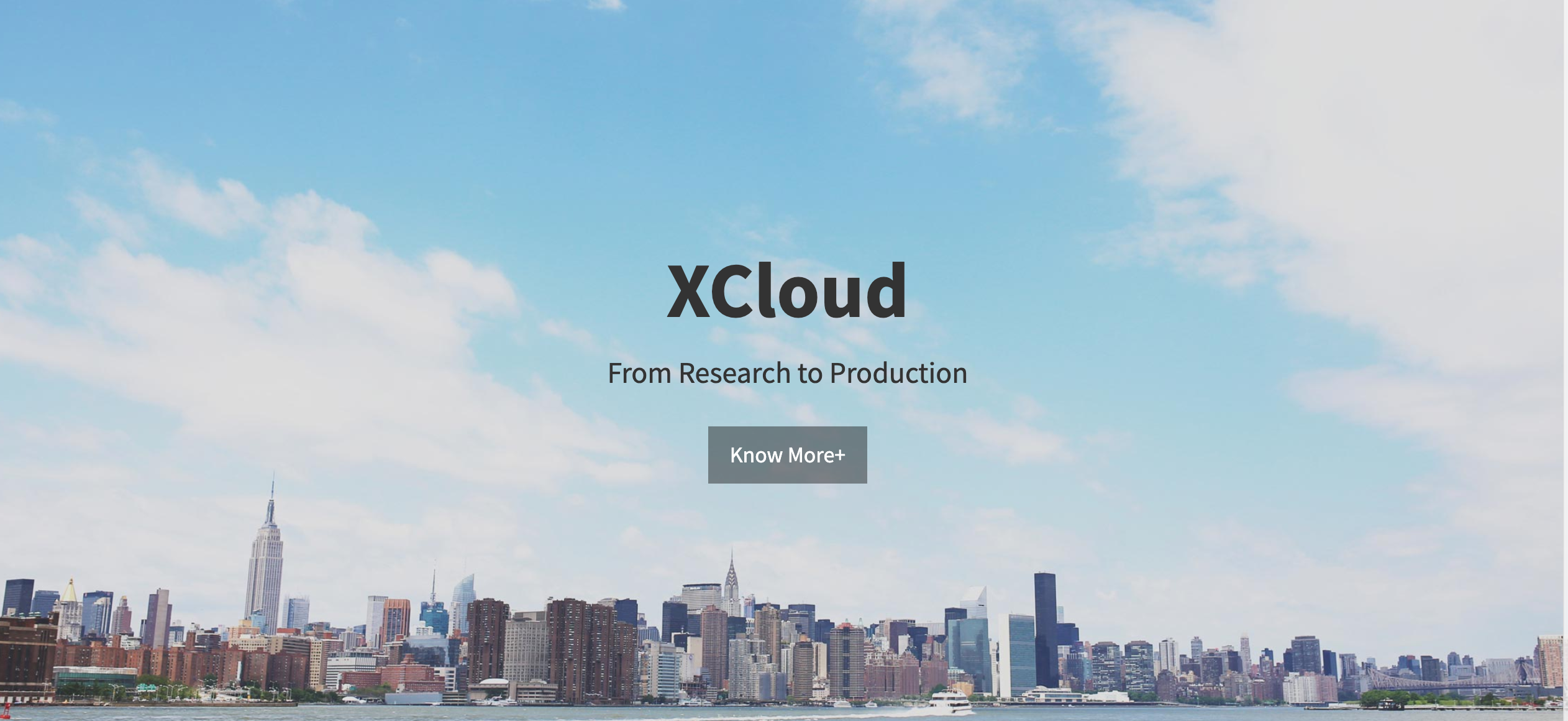}
\end{figure}

From Table~\ref{tab:test} we can conclude that the performance and stability of \textit{XCloud} are quite satisfactory under current software and hardware condition. We believe the performance could be heavily improved if stronger hardware is provided. The test environment with 2080TI GPUs and Intel XEON CPU is enough to support 20 QPS (query per second). By deploying \textit{XCloud} on your machine and running server, you will get the homepage as Figure~\ref{fig:index}.

\section{Conclusion and Future Work}
\label{sec:conclusion}
In this paper, we construct an AI cloud platform with high performance and stability which provides common AI service in form of RESTful API, to ease the development of AI projects. In our future work, we will integrate more service into \textit{XCloud} and develop better models with advanced performance.

\bibliographystyle{plain}
\bibliography{sample-base}

\begin{thebibliography}{10}

\bibitem{amodei2016deep}
Dario Amodei, Sundaram Ananthanarayanan, Rishita Anubhai, Jingliang Bai, Eric
  Battenberg, Carl Case, Jared Casper, Bryan Catanzaro, Qiang Cheng, Guoliang
  Chen, et~al.
\newblock Deep speech 2: End-to-end speech recognition in english and mandarin.
\newblock In {\em International conference on machine learning}, pages
  173--182, 2016.

\bibitem{bosse2016deep}
Sebastian Bosse, Dominique Maniry, Thomas Wiegand, and Wojciech Samek.
\newblock A deep neural network for image quality assessment.
\newblock In {\em 2016 IEEE International Conference on Image Processing
  (ICIP)}, pages 3773--3777. IEEE, 2016.

\bibitem{esteva2017dermatologist}
Andre Esteva, Brett Kuprel, Roberto~A Novoa, Justin Ko, Susan~M Swetter,
  Helen~M Blau, and Sebastian Thrun.
\newblock Dermatologist-level classification of skin cancer with deep neural
  networks.
\newblock {\em Nature}, 542(7639):115, 2017.

\bibitem{he2015delving}
Kaiming He, Xiangyu Zhang, Shaoqing Ren, and Jian Sun.
\newblock Delving deep into rectifiers: Surpassing human-level performance on
  imagenet classification.
\newblock In {\em Proceedings of the IEEE international conference on computer
  vision}, pages 1026--1034, 2015.

\bibitem{he2016deep}
Kaiming He, Xiangyu Zhang, Shaoqing Ren, and Jian Sun.
\newblock Deep residual learning for image recognition.
\newblock In {\em Proceedings of the IEEE conference on computer vision and
  pattern recognition}, pages 770--778, 2016.

\bibitem{huang2017densely}
Gao Huang, Zhuang Liu, Laurens Van Der~Maaten, and Kilian~Q Weinberger.
\newblock Densely connected convolutional networks.
\newblock In {\em Proceedings of the IEEE conference on computer vision and
  pattern recognition}, pages 4700--4708, 2017.

\bibitem{johnson2017google}
Melvin Johnson, Mike Schuster, Quoc~V Le, Maxim Krikun, Yonghui Wu, Zhifeng
  Chen, Nikhil Thorat, Fernanda Vi{\'e}gas, Martin Wattenberg, Greg Corrado,
  et~al.
\newblock Google’s multilingual neural machine translation system: Enabling
  zero-shot translation.
\newblock {\em Transactions of the Association for Computational Linguistics},
  5:339--351, 2017.

\bibitem{kang2014convolutional}
Le~Kang, Peng Ye, Yi~Li, and David Doermann.
\newblock Convolutional neural networks for no-reference image quality
  assessment.
\newblock In {\em Proceedings of the IEEE conference on computer vision and
  pattern recognition}, pages 1733--1740, 2014.

\bibitem{kang2015simultaneous}
Le~Kang, Peng Ye, Yi~Li, and David Doermann.
\newblock Simultaneous estimation of image quality and distortion via
  multi-task convolutional neural networks.
\newblock In {\em 2015 IEEE international conference on image processing
  (ICIP)}, pages 2791--2795. IEEE, 2015.

\bibitem{krizhevsky2012imagenet}
Alex Krizhevsky, Ilya Sutskever, and Geoffrey~E Hinton.
\newblock Imagenet classification with deep convolutional neural networks.
\newblock In {\em Advances in neural information processing systems}, pages
  1097--1105, 2012.

\bibitem{lecun2015deep}
Yann LeCun, Yoshua Bengio, and Geoffrey Hinton.
\newblock Deep learning.
\newblock {\em nature}, 521(7553):436, 2015.

\bibitem{leff2001web}
Avraham Leff and James~T Rayfield.
\newblock Web-application development using the model/view/controller design
  pattern.
\newblock In {\em Proceedings fifth ieee international enterprise distributed
  object computing conference}, pages 118--127. IEEE, 2001.

\bibitem{liang2018scut}
Lingyu Liang, Luojun Lin, Lianwen Jin, Duorui Xie, and Mengru Li.
\newblock Scut-fbp5500: A diverse benchmark dataset for multi-paradigm facial
  beauty prediction.
\newblock In {\em 2018 24th International Conference on Pattern Recognition
  (ICPR)}, pages 1598--1603. IEEE, 2018.

\bibitem{Liu_2017_CVPR}
Weiyang Liu, Yandong Wen, Zhiding Yu, Ming Li, Bhiksha Raj, and Le~Song.
\newblock Sphereface: Deep hypersphere embedding for face recognition.
\newblock In {\em The IEEE Conference on Computer Vision and Pattern
  Recognition (CVPR)}, 2017.

\bibitem{paszke2019pytorch}
Adam Paszke, Sam Gross, Francisco Massa, Adam Lerer, James Bradbury, Gregory
  Chanan, Trevor Killeen, Zeming Lin, Natalia Gimelshein, Luca Antiga, et~al.
\newblock Pytorch: An imperative style, high-performance deep learning library.
\newblock In {\em Advances in Neural Information Processing Systems}, pages
  8024--8035, 2019.

\bibitem{sculley2015hidden}
David Sculley, Gary Holt, Daniel Golovin, Eugene Davydov, Todd Phillips,
  Dietmar Ebner, Vinay Chaudhary, Michael Young, Jean-Francois Crespo, and Dan
  Dennison.
\newblock Hidden technical debt in machine learning systems.
\newblock In {\em Advances in neural information processing systems}, pages
  2503--2511, 2015.

\bibitem{silver2016mastering}
David Silver, Aja Huang, Chris~J Maddison, Arthur Guez, Laurent Sifre, George
  Van Den~Driessche, Julian Schrittwieser, Ioannis Antonoglou, Veda
  Panneershelvam, Marc Lanctot, et~al.
\newblock Mastering the game of go with deep neural networks and tree search.
\newblock {\em nature}, 529(7587):484, 2016.

\bibitem{sun2016benchmark}
Xiaoxiao Sun, Jufeng Yang, Ming Sun, and Kai Wang.
\newblock A benchmark for automatic visual classification of clinical skin
  disease images.
\newblock In {\em European Conference on Computer Vision}, pages 206--222.
  Springer, 2016.

\bibitem{talebi2018nima}
Hossein Talebi and Peyman Milanfar.
\newblock Nima: Neural image assessment.
\newblock {\em IEEE Transactions on Image Processing}, 27(8):3998--4011, 2018.

\bibitem{wen2016discriminative}
Yandong Wen, Kaipeng Zhang, Zhifeng Li, and Yu~Qiao.
\newblock A discriminative feature learning approach for deep face recognition.
\newblock In {\em European conference on computer vision}, pages 499--515.
  Springer, 2016.

\bibitem{wu2019ip102}
Xiaoping Wu, Chi Zhan, Yu-Kun Lai, Ming-Ming Cheng, and Jufeng Yang.
\newblock Ip102: A large-scale benchmark dataset for insect pest recognition.
\newblock In {\em Proceedings of the IEEE Conference on Computer Vision and
  Pattern Recognition}, pages 8787--8796, 2019.

\bibitem{xu2019hierarchical}
Lu~Xu, Heng Fan, and Jinhai Xiang.
\newblock Hierarchical multi-task network for race, gender and facial
  attractiveness recognition.
\newblock In {\em 2019 IEEE International Conference on Image Processing
  (ICIP)}, pages 3861--3865. IEEE, 2019.

\bibitem{xu2019data}
Lu~Xu, Jinhai Xiang, Yating Wang, and Fuchuan Ni.
\newblock Data-driven approach for quality evaluation on knowledge sharing
  platform.
\newblock {\em arXiv preprint arXiv:1903.00384}, 2019.

\bibitem{xu2018crnet}
Lu~Xu, Jinhai Xiang, and Xiaohui Yuan.
\newblock Crnet: Classification and regression neural network for facial beauty
  prediction.
\newblock In {\em Pacific Rim Conference on Multimedia}, pages 661--671.
  Springer, 2018.

\bibitem{yang2017visual}
Fan Yang, Ajinkya Kale, Yury Bubnov, Leon Stein, Qiaosong Wang, Hadi Kiapour,
  and Robinson Piramuthu.
\newblock Visual search at ebay.
\newblock In {\em Proceedings of the 23rd ACM SIGKDD International Conference
  on Knowledge Discovery and Data Mining}, pages 2101--2110. ACM, 2017.

\bibitem{zhang2018visual}
Yanhao Zhang, Pan Pan, Yun Zheng, Kang Zhao, Yingya Zhang, Xiaofeng Ren, and
  Rong Jin.
\newblock Visual search at alibaba.
\newblock In {\em Proceedings of the 24th ACM SIGKDD International Conference
  on Knowledge Discovery \& Data Mining}, pages 993--1001. ACM, 2018.

\end{thebibliography}
\end{document}